\documentclass[aps,prb,twocolumn,superscriptaddress,
groupedaddress,10pt]{revtex4}
\usepackage{graphicx}
\usepackage{bm}
\usepackage{amsmath}
\usepackage{amssymb}
\usepackage{amsfonts}
\usepackage{xcolor}
\usepackage{listings}
\usepackage{matlab-prettifier}
\usepackage{dsfont}
\usepackage{braket}
\newcommand\re{{\rm e}}
\newcommand\dif[2]{\frac{{{\rm d} #1}}{{\rm d}#2}}
\newcommand\di{{\rm d}}
\newcommand\ri{{\rm i}}
\newcommand\partd[2]{\frac{\partial #1}{\partial #2}}
\newcommand\Ws{WSe$_2$ }

\begin{document}
	
	\title{Theoretical study of spin-dependent transport in WSe$_2$-based vertical spin valves}
	
	\author{Yibo Wang}
	\affiliation{School of Physics, Nankai University, Tianjin 300071, China}
	
	\author{Yuchen Liu}
	\affiliation{School of Physics, Nankai University, Tianjin 300071, China}
	\affiliation{Department of Materials Engineering Science, The University of Osaka, Toyonaka, Osaka 560-8531, Japan}
	
	\author{Xinhe Wang}
	\email{xinhe@buaa.edu.cn}
	\affiliation{Fert Beijing Institute, School of Integrated Circuit Science and Engineering, Beihang University, Beijing, China.}
	\affiliation{State Key Laboratory of Spintronics, Hangzhou International Innovation Institute, Beihang University, Hangzhou, China.}
	
    \author{Wang Yang}
    \email{wyang@nankai.edu.cn}
    \affiliation{School of Physics, Nankai University, Tianjin 300071, China}

    \begin{abstract}
We theoretically investigate spin-dependent transport in a TMD-based vertical spin valve, taking WSe$_2$ as a representative example. 
Using effective Hamiltonians for the heterostructure and the Landauer formula, we derive the transmission and reflection coefficients within a transfer-matrix approach. 
The calculated magnetoresistance shows an oscillatory dependence on the WSe$_2$ thickness when the Fermi level is tuned near the valence-band maximum. 
The effects of gate voltage and exchange fields on the magnetoresistance are further analyzed. 
We also identify a Fabry-P\'erot-like interference contribution to the magnetoresistance, which can enhance or even induce negative magnetoresistance in certain thickness regimes.
Our results provide a qualitative understanding of the negative magnetoresistance observed in WSe$_2$-based spin valves and may offer useful insights for the design of tunable spintronic devices.
    \end{abstract}
    \maketitle
	\section{Introduction}
	\label{sec:intro}
	
Two-dimensional (2D) materials are characterized by their atomic thickness and tunable electronic structures \cite{novoselov2004electric,geim2007rise,mak2010atomically,novoselov2012roadmap,dragoman20162d}. Unlike bulk crystals, their reduced dimensionality gives rise to quantum confinement, enabling control over charge, spin, and valley degrees of freedom \cite{macneill2015breaking,xu2014spin,dean2013hofstadter}. The van der Waals nature of these materials facilitates the assembly of heterostructures with relatively clean interfaces, providing a useful platform for nanoelectronic and spintronic devices \cite{britnell2012field,schaibley2016valleytronics,liu2018approaching}. Furthermore, the sensitivity of 2D systems to external fields and doping offers a versatile setting for tuning physical properties and exploring novel quantum phenomena \cite{manzeli20172d,frisenda2018atomically}.

Within this family, transition metal dichalcogenides (TMDs) such as $\mathrm{MoS}_2$, $\mathrm{WS}_2$, and $\mathrm{WSe}_2$ are distinguished by their sizable energy gaps and substantial spin-orbit coupling (SOC)~\cite{xiao2012coupled,aivazian2015magnetic,wang2018colloquium,fang2015ab,rostami2013effective}, making them suitable systems for studying spin-dependent transport. In particular, the heavy tungsten atoms make $\mathrm{WSe}_2$ one of the TMDs with relatively strong SOC (approximately $0.5$\,eV at the valence band $\rm K$ point)~\cite{zhang2016electronic}, which gives rise to spin--valley locking and valley-contrasting optical selection rules. The low-energy physics of these systems can be effectively described by a $\bm{k}\cdot\bm{p}$ Hamiltonian, where the valley degree of freedom serves as a ``pseudospin'' analogous to the sublattice pseudospin in graphene~\cite{xiao2012coupled,de2017strong}. These electronic properties make $\mathrm{WSe}_2$ a promising candidate for realizing spin-filtering and spin-valve effects in van der Waals heterostructures, with potential relevance for spintronic devices~\cite{zheng2022spin,zambrano2021spin,song2023van}.

While research on TMDs and other 2D materials has predominantly focused on in-plane carrier transport~\cite{zhai2008theory,zou2009negative,roy20162d,hajati2021spin}, vertical transport -- perpendicular to the atomic layers -- is also important for understanding electronic behavior and exploring device geometries. Unlike the high-conductivity in-plane channels, cross-plane transport is governed by distinct mechanisms, including interlayer coupling, interfacial band alignment, and tunneling barrier heights. In TMD systems, charge transfer between layers and across electrode interfaces gives rise to a range of transport phenomena~\cite{zhu2017vertical,golsanamlou2022vertical,rosul2019thermionic,song2018giant}. 
	
In the work by Wang \textit{et al.} reported in Ref. \onlinecite{wang2022spin}, a WSe$_2$-based spin-valve device was fabricated, revealing a giant valley-Zeeman-type spin-orbit field (SOF) \cite{wang2022spin}. Under this mechanism, carrier spins can be manipulated or even reversed by tuning the gate voltage or the number of WSe$_2$ layers, without the need for large external magnetic fields or high driving currents, thereby indicating the potential of this platform for low-power spintronic applications. Moreover, the magnetoresistance in such devices was observed to oscillate in sign as a function of the WSe$_2$ thickness, consistent with earlier studies \cite{zhao2017magnetic}. A phenomenological model was further proposed to account for this behavior.
	
In this work, we theoretically analyze spin-dependent vertical transport in a vertical spin valve composed of ferromagnetically doped graphene and WSe$_2$ middle layers. Within the Landauer framework, we compute the magnetoresistance ratio as functions of thickness, gate potential, and electrode exchange fields, and find an oscillatory magnetoresistance as a function of WSe$_2$ thickness that is most pronounced when the Fermi level in the WSe$_2$ region is tuned near the valence band maximum. These results provide a qualitative understanding of the negative magnetoresistance observed in related devices\cite{wang2022spin,zhao2017magnetic} and may offer useful insights for engineering tunable spintronic devices in TMD-based heterostructures. \par

In addition to the spin-dependent transport mechanism related to SOC,
we also identify a Fabry-Pérot-like interference contribution to the magnetoresistance,
which can lead to negative magnetoresistance even in the absence of SOC and an effective Zeeman field in the inter-layer material. 
More precisely, the interference effects arising from multiple reflections and transmissions can be either constructive or destructive depending on the thickness of the inter-layer spacer region. 
For certain thicknesses at which destructive interference occurs, the anti-parallel configuration can suppress the interference effects,
thereby reducing the destructive effect relative to the parallel configuration. 
As a result, the tunneling conductance in the anti-parallel configuration can become larger than that in the parallel configuration,
leading to negative magnetoresistance even without SOC.

The remainder of this paper is organized as follows. In Sec. \ref{sec:ham}, we introduce the model Hamiltonians for the different regions of the heterostructure. In Sec. \ref{sec:trans}, we formulate the transfer-matrix approach for spin-dependent transport and derive the transmission coefficients. Section \ref{sec:magneto} is devoted to the evaluation of the magnetoresistance ratio within the Landauer formalism, together with a discussion of the density of states and spin polarization in the electrodes. In Sec. \ref{sec:numer}, we present a detailed analysis of the magnetoresistance as functions of the spacer thickness, gate potential, and exchange field, and discuss the corresponding MR characteristics. Section \ref{subsec:interference_MR} further examines the contribution of Fabry-Pérot-like interference to the magnetoresistance. Finally, Sec. \ref{sec:summary} summarizes the main results and the physical insights into WSe$_2$-based vertical spin valves.

\section{Model Hamiltonian}

We consider a vertical spin valve composed of three regions, where regions I and III correspond to the bottom and top electrodes, respectively, and region II is the intermediate multi-layer WSe$_2$ spacer, as schematically illustrated in Fig.~\ref{fig:verticalspinvalve}. Although realistic spin-transport devices usually employ ferromagnetic metallic electrodes, we model the electrodes here as doped thick graphene layers in external magnetic fields \cite{karpan2007graphite}. This modeling provides a tractable theoretical framework and allows us to focus on the essential physics of spin-dependent vertical transport.

The above description is, however, an idealization of realistic devices. In particular, interface-related effects such as Schottky barriers, band alignment and Fermi-level pinning, charge transfer, and other microscopic interface details are not explicitly included. A fully quantitative treatment of experiments would require these ingredients. Nevertheless, for the qualitative understanding of the negative magnetoresistance and its oscillatory behavior with WSe$_2$ thickness, the present effective model is sufficient.

\subsection{Hamiltonians}	
\label{sec:ham}



\begin{figure}
		\centering
		\includegraphics[width=7cm]{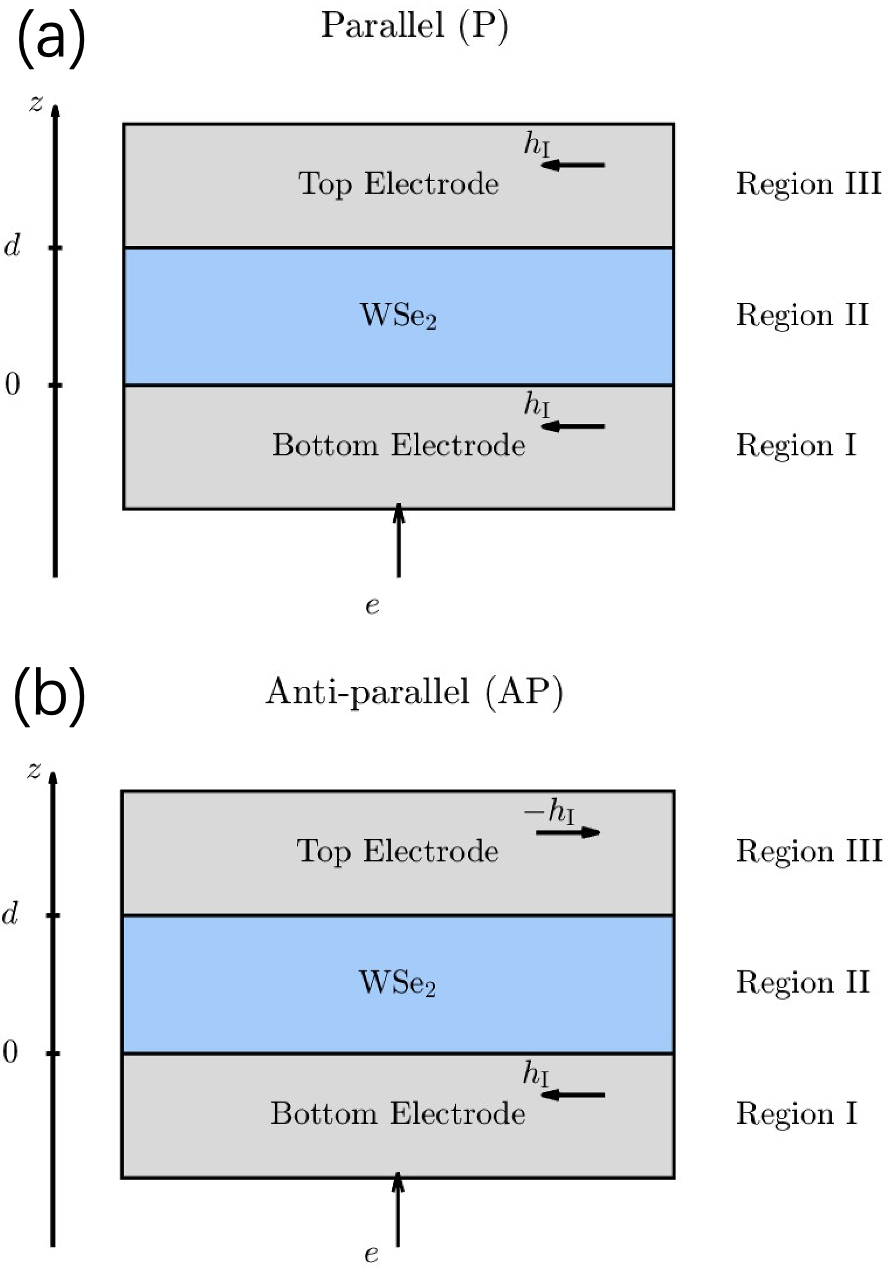}
		\caption{Schematic illustrations of the vertical spin valve based on WSe$_2$ spacer for (a) parallel and (b) anti-parallel configurations. The device consists of three regions: bottom electrode (region I, $z<0$), central WSe$_2$ spacer (region II, $0<z<d$), and top electrode (region III, $z>d$). Exchange fields in the electrodes are represented by arrows: in the parallel configuration (P) both electrodes have magnetization along the same direction; in the anti-parallel configuration (AP), the magnetizations are opposite.}
		\label{fig:verticalspinvalve}
\end{figure}

With the aforementioned modeling, the Hamiltonian of region I is given by \cite{castro2009electronic}
\begin{equation}
	H_{\rm I}=v_0(k_x\sigma_x+k_y\sigma_y)-h_{\rm I}S_x-\frac{\hbar^2}{2m_0}\partial_z^2-\varepsilon_F,
	\label{HamI}
\end{equation}
in which $v_0$ is the Fermi velocity in the thick graphene layers near the K point; 
the term $-h_{\rm I}S_x$ represents a Zeeman field along the $x$ direction and introduces ferromagnetic spin polarization in the electrodes; 
$m_0$ is the effective mass along the vertical direction; 
$\sigma_\alpha$ ($\alpha=x,y,z$) are the Pauli matrices acting on the sublattice pseudospin space associated with the two graphene sublattices $A$ and $B$, while $S_\alpha$ are the Pauli matrices in the real spin space. 
The Hamiltonian in region III can be written as
\begin{equation}
	H_{\rm III}=v_0(k_x\sigma_x+k_y\sigma_y)-h_1S_x-h_2S_y-\frac{\hbar^2}{2m_0}\partial_z^2-\varepsilon_F,
	\label{HamIII}
\end{equation}
which differs from that in region I only by the direction of the exchange field.
	
\begin{figure}
		\centering
		\includegraphics[width=8.5cm]{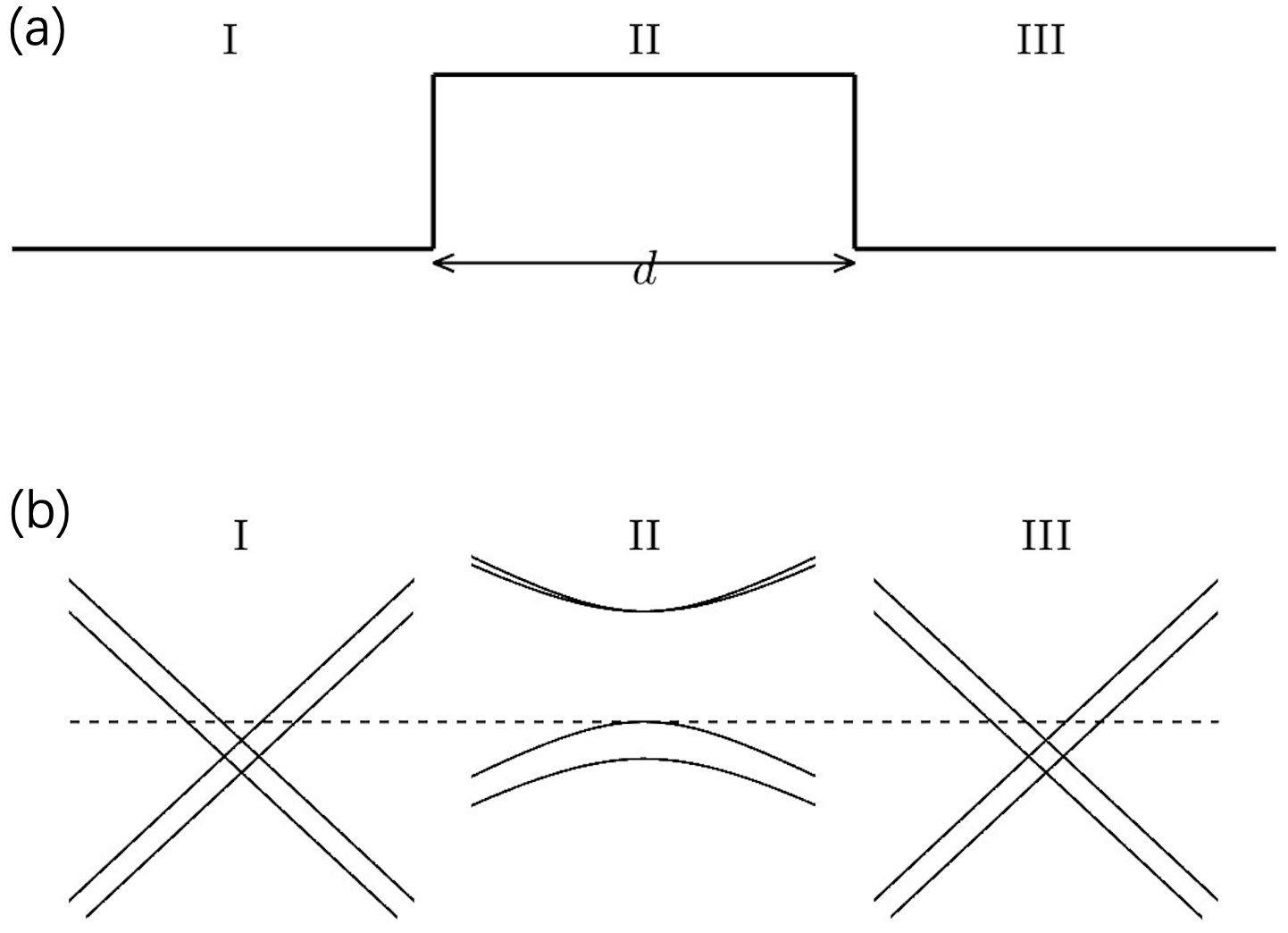}
		\caption{(a) Potential profile along the vertical transport direction, and (b) The in-plane band structures in the three regions. Note that in this figure the three regions are arranged horizontally for clarity,
which is rotated $90^\circ$ with respect to Fig. \ref{fig:verticalspinvalve}.		
The horizontal arrangement is adopted to better illustrate the band alignment and Fermi level matching.}
		\label{fig:bandstruct}
\end{figure}

Region II is the central layer through which carriers transfer from the bottom to the top electrodes. 
The primary example of region II under consideration is multi-layer \Ws,
though other TMD materials can be considered in a similar manner by adjusting the parameters in region II accordingly. 
The effective Hamiltonian in region II is given by \cite{xiao2012coupled,wang2022spin,kormanyos2015kp}
\begin{equation}
H_{\rm II}=v(k_x\sigma_x+k_y\sigma_y)+\frac\Delta2\sigma_z-\lambda\frac{\sigma_z-1}{2}S_z-\frac{\hbar^2}{2m}\partial_z^2+V_g, 
\label{HamII}
\end{equation}
in which $v$ is the velocity parameter of \Ws  near the K point; 
$\Delta$ is the direct band gap at ${\rm K}$ point;
$\lambda$ is the strength of the effective Zeeman field which causes spin splitting of the valence band; 
$m$ is the effective mass along the vertical direction;
and $V_g$ is an external gate potential. 
A schematic plot of the band structures of the three regions is shown in Fig. \ref{fig:bandstruct},
where the geometry is rotated by $90^\circ$ with respect to Fig. \ref{fig:verticalspinvalve} so that the alignment of Fermi levels in different regions is more transparent.

It is worth to note that 
if the spin magnetizations in the bottom and top electrodes (i.e., regions I and III) are assumed to be either parallel or anti-parallel with each other,
then only considering the $\rm K$ valley is enough, since the ${\rm K}^\prime$ valley gives identical contribution to the magnetoresistance, which is enforced by symmetry constraints as follows. 
In the absence of exchange fields (i.e., $h_{\text{I}}=h_1=h_2=0$),
the whole system including both ${\rm K}$ and ${\rm K}^\prime$ valleys is invariant under the spin-orbit coupled mirror reflection operation $M_x$, which sends $x$ to $-x$, keeping $y$ unchanged. 
Now consider the case in which $h_{\rm I}$ and $h_1$ are nonzero while $h_2=0$. 
In this situation, the spin polarizations in both region I and region III are along the $x$ direction, corresponding to the parallel or anti-parallel configurations of the electrode magnetizations.
Notice that in such case,  $M_x$ remains to be a symmetry of regions I, II, and III, since spin operators form a pseudo-vector and $S_x$ does not change under $M_x$. 
As a result, the tunneling conductances from ${\rm K}$ and ${\rm K}^\prime$ valleys are related by the mirror reflection operation $M_x$. 
Since $M_x$ does not change $S_x$,
the magnetoresistance -- defined as the difference between tunneling conductances of parallel ($h_1$ having the same sign as $h_{\text{I}}$) and anti-parallel ($h_1$ having the opposite sign as $h_{\text{I}}$) cases -- must be equal for the ${\rm K}$ and ${\rm K}^\prime$ valleys. 
We emphasize that the equality between the contributions from ${\rm K}$ and ${\rm K}^\prime$ valleys does not hold for more general cases when $h_2\neq 0$ in Eq. (\ref{HamIII}).

It is worth to note that the modeling adopted in this work implicitly assumes an AA-stacked configuration of layers of WSe$_2$. 
For AA-stacking and nearest-neighboring inter-layer hopping,
the out-of-plane dispersion is given by $E(k_z)= -t_\perp\cos(k_z d_0)$, where 
$k_z$ is the wave-vector along $z$-direction and
$d_0$ is the distance between two adjacent layers. 
For low energy transport, $k_z$ is small
and the out-of-plane dispersion can be approximated as a parabolic form with effective mass $m^\ast=\frac{\hbar^2}{t_\perp d^2}$. 
This justifies the continuum version of the Hamiltonian used in Eq. (\ref{HamII}). 
In contrast, for other stacking configurations such as AB stacking, the interlayer coupling is more complicated,  
leading to an entangled energy dispersion between in-plane and out-of-plane momenta that cannot be decomposed into a simple sum of two parts, each involving  only the in-plane momentum or the out-of-plane one. 
As a result, an effective mass approximation cannot be used to describe vertical spin transport in AB or other stacking configurations,
which requires a separate treatment and will be left for future investigations. 

\subsection{Model parameters}
\label{subsec:model_parameters}

Notice that Eqs. (\ref{HamI},\ref{HamIII},\ref{HamII}) are effective Hamiltonians which are valid only at low energies. 
The applicable range of these low energy Hamiltonians is characterized by a cutoff $k_c$ in the in-plane momentum space,
which is part of the definition of the low energy Hamiltonians. 
Furthermore, the Fermi energy $\epsilon_F$ is also a parameter that has to be fed into the model to carry out any concrete calculation. 

The two parameters $k_c$ and $\epsilon_F$ should be determined by fitting the experimental results.
We use two experimental quantities to fit $k_c$ and $\epsilon_F$:
One is the magnetization in the ferromagnetic electrodes,
and the other is the magnetoresistance for monolayer WSe$_2$. 
Detailed fitting method will be discussed in Sec. \ref{subsec:detail_fitting}.
Once $k_c$ and $\epsilon_F$ are determined, they can be used to calculate magnetoresistance under other conditions,
especially when the layer number, gating potential $V_g$, and doping are changed.

\section{Transfer matrix approach to spin-dependent transport}
\label{sec:trans}
    
In this section, we investigate the spin-dependent tunneling properties of the vertical spin valve composed of multi-layer \Ws\ and the electrodes within a scattering approach. For given in-plane momentum $(k_x,k_y)$ and energy $E$, the scattering wavefunctions in each region shown in Fig.~\ref{fig:bandstruct}(a) are obtained from the Hamiltonians in Eqs.~(\ref{HamI}, \ref{HamIII}, \ref{HamII}) and then matched across the interfaces between adjacent regions through appropriate boundary conditions.    

\subsection{Wavefunctions in regions I, II, III}

In region $\Gamma$ ($\Gamma=\text{I,~II,~III}$) and for given in-plane momentum $(k_x,k_y)$ and energy $E$,
there are two values of $k_z$ that satisfies the eigen-equation of the Hamiltonian $H_{\Gamma}$ in Eqs. (\ref{HamI},\ref{HamIII},\ref{HamII}),
one left-moving and the other right-moving which differ by an overall sign.   
For each $k_z$, there are four eigen-wavefunctions since the system is four-component. 
As a result, the wavefunction in  region $\Gamma$ can be written as a sum of eight terms 
\begin{equation}
    	\Psi^{(\Gamma)}(z)=\sum\limits_{i=1}^4\big[A_i^{(\Gamma)}\re^{\ri k_{z,i}^{(\Gamma)}z}+B_i^{(\Gamma)}\re^{-\ri k_{z,i}^{(\Gamma)}z}\big]\Phi_i^{(\Gamma)}, 
    	\label{wavef}
\end{equation}
in which $A_i^{(\Gamma)}$ and $B_i^{(\Gamma)}$ are the coefficients for the right-moving and left-moving components, respectively,
and $\Phi_i^{(\Gamma)}$ is a four-component spinor to be determined later. 

For later convenience, we introduce the following notations for spinors. 
For any vector $\bm{u}$, let $\Xi_\alpha(\bm{u})$ denote the eigenstate of $\bm{u}\cdot\bm{\sigma}$ with eigenvalue $\alpha|\bm{u}|$, where     $\alpha=\pm 1$ and $\bm{\sigma}=(\sigma_x,\sigma_y,\sigma_z)^T$.
More explicitly, if $\bm{u}=|\bm{u}|(\sin\theta\cos\phi,\sin\theta\sin\phi,\cos\theta)^T$, then 
\begin{eqnarray}
\Xi_+(\bm{u})&=&\begin{pmatrix}
    	\cos\frac{\theta}{2}\re^{-\ri\phi}\\
    	\sin\frac{\theta}{2}
    	\end{pmatrix},\nonumber\\ 
\Xi_-(\bm{u})&=&\begin{pmatrix}
    	-\sin\frac{\theta}{2}\re^{-\ri\phi}\\
    	\cos\frac{\theta}{2}
    	\end{pmatrix}. 
\end{eqnarray}
    Similarly, for any vector $\bm{v}$, let $\chi_\alpha(\bm{v})$ denote the eigenstate of $\bm{v}\cdot\bm{S}$ corresponding to eigenvalue $\alpha|\bm{v}|$,
    where $\alpha=\pm 1$ and $\bm{S}=(S_x,S_y,S_z)^T$.
For example, $\chi_+(\hat{\bm{z}})=(1,0)^T,\chi_-(\hat{\bm{z}})=(0,1)^T$.

\subsubsection{Regions I and III}

We first solve the wavefunctions in regions I. 
Define $H_{\rm I}^\prime$ as 
\begin{equation}
    	H_{\rm I}^\prime=v_0(k_x\sigma_x+k_y\sigma_y)-h_{\rm I}S_x. 
\end{equation}
The eigenvalues of $H_I^\prime$ can be solved as 
\begin{equation}
    	\mathcal{E}_i^{(\rm I)}=\alpha_iv_0k_{\parallel}+\beta_i h_{\rm I}, 
    	\label{eigenI}
\end{equation}
    where $k_\parallel=\sqrt{k_x^2+k_y^2}$ and \begin{equation}
    	(\alpha_i,\beta_i)=\begin{cases}
    		(+1,-1),i=1\\
    		(+1,+1),i=2\\
    		(-1,-1),i=3\\
    		(-1,+1),i=4
    	\end{cases}. 
	\label{eq:alpha_beta_i}
    \end{equation}	    
For given energy $E$,
there are eight solutions $\pm k_{z,i}^{(\text{I})}$ of the momentum along $z$-direction, given by 
\begin{equation}
    	k_{z,i}^{(\rm I)}=\sqrt{\frac{2m_0}{\hbar^2}(E+\varepsilon_F-\mathcal{E}_i^{(\rm I)})}. 
\end{equation}

The two spinors which are eigenfunctions of $k_x\sigma_x+k_y\sigma_y$ can be solved explicitly as 
\begin{eqnarray}
    	\Xi_{+}(\bm{k}_\parallel)&=&\frac{1}{\sqrt2}\begin{pmatrix}
    		\frac{k_x-\ri k_y}{k_\parallel}\\
    		1
    	\end{pmatrix}\nonumber\\
	\Xi_{-}(\bm{k}_\parallel)&=&\frac{1}{\sqrt2}\begin{pmatrix}
    		\frac{k_x-\ri k_y}{k_\parallel}\\
    		-1
    	\end{pmatrix}, 
\end{eqnarray}
    where  $\bm{k}_\parallel=(k_x,k_y,0)^T$ is the in-plane momentum and $k_\parallel=\sqrt{k_x^2+k_y^2}$.
    The eigenfunctions of $S_x$ are
\begin{eqnarray}
    	\chi_{+}(\hat{\bm{x}})&=&\frac{1}{\sqrt2}\begin{pmatrix}
    		1\\1
    	\end{pmatrix} \nonumber\\
    	\chi_{-}(\hat{\bm{x}})&=&\frac{1}{\sqrt2}\begin{pmatrix}
    		1\\-1
    	\end{pmatrix}. 
\end{eqnarray}
    Therefore, the four four-component spinors $\Phi_i^{(\rm I)}$ in region I can be solved as
\begin{eqnarray}
    	\Phi_i^{(\rm I)}=\Xi_{\alpha_i}({\bm{k}}_\parallel)\otimes\chi_{\beta_i}(\hat{\bm{x}}). 
    	\label{wfI}
\end{eqnarray}

Region III can be treated in a similar manner as region I. 
Define $H_{\rm III}^\prime$ as 
\begin{equation}
    	H_{\rm III}^\prime=v_0(k_x\sigma_x+k_y\sigma_y)-h_1S_x-h_2S_y. 
    \end{equation}
    The eigenvalues of $H_{\rm III}^\prime$ can be solved as
    \begin{equation}
    	\mathcal{E}_i^{\rm (III)}=\alpha_i v_0k_{\parallel}+\beta_i h_\parallel, 
\end{equation}
    where $h_\parallel=\sqrt{h_1^2+h_2^2}$, and $(\alpha_i,\beta_i)$ are defined in Eq. (\ref{eq:alpha_beta_i}).  
The wavevectors along $z$-direction for a given energy $E$ are given by
    \begin{equation}
    	k_{z,i}^{(\rm III)}=\sqrt{\frac{2m_0}{\hbar^2}(E+\varepsilon_F-\mathcal{E}_i^{(\rm III)})}. 
    \end{equation}
The eigenfunctions of $h_1S_x+h_2S_y$ are
\begin{eqnarray}
\chi_{+}(\bm{h})&=&\frac{1}{\sqrt2}\begin{pmatrix}
    		\frac{h_1-\ri h_2}{h_\parallel}\\
    		1
    	\end{pmatrix}\nonumber\\
\chi_{-}(\bm{h})&=&\frac{1}{\sqrt2}\begin{pmatrix}
    		\frac{h_1-\ri h_2}{h_\parallel}\\
    		-1
    	\end{pmatrix}, 
\end{eqnarray}
    where $\bm{h}=(h_1,h_2,0)^T$.
Therefore, the four four-component spinors $\Phi_i^{(\rm III)}$ in region III can be written as 
    \begin{equation}
    	\Phi_i^{(\rm III)}=\Xi_{\alpha_i}(\bm{k}_\parallel)\otimes\chi_{\beta_i}(\bm{h}). 
    \end{equation}
    
\subsubsection{Region II}
    
Next we solve the wavefunctions in regions II. 
Define $H_{\rm II}^\prime$ as 
    \begin{equation}
    	H_{\rm II}^\prime=v(k_x\sigma_x+k_y\sigma_y)+\frac{\Delta}{2}\sigma_z-\lambda\frac{\sigma_z-1}{2}S_z. 
    \end{equation}
Since $S_z$ is a good quantum number, $H_{\rm II}^\prime$ can be block-diagonalized into two $2\times 2$ sectors labeled by $S_z=\beta$ with $\beta=\pm1$. The corresponding block Hamiltonian is $\bm{U}_\beta\cdot\bm{\sigma}+\frac{1}{2}\beta\lambda$, where $\bm{U}_\beta=(vk_x,vk_y,\frac{1}{2}(\Delta-\beta\lambda))$. Its eigenvalues are $\alpha|\bm{U}_\beta|+\frac{1}{2}\beta\lambda$ with $\alpha=\pm1$.

The four eigenvalues of $H_{\text{II}}^\prime$  can be solved as 
\begin{equation}
    	\mathcal{E}_i^{(\rm II)}=\alpha_i\sqrt{v^2k_{\parallel}^2+\frac14(\Delta-\beta_i\lambda)^2}+\frac12\beta_i\lambda,
    \end{equation}
where $i\in\{1,2,3,4\}$, and the values of $(\alpha_i,\beta_i)$ are the same as Eq. (\ref{eq:alpha_beta_i}).
The momenta $k_{z,i}^{(\rm II)}$ along the vertical $z$-direction in region II are solved as
    \begin{equation}
    	k_{z,i}^{(\rm II)}=\sqrt{\frac{2m}{\hbar^2}(E-V_g-\mathcal{E}_i^{(\rm II)})}.
    \end{equation}
    The four four-component spinors  $\Phi_i^{(\rm II)}$ in region II can be expressed as
    \begin{equation}
    	\Phi_i^{(\rm II)}=\Xi_{\alpha_i}(\bm{U}_{\beta_i})\otimes\chi_{\beta_i}(\hat{\bm{z}}). 
    \end{equation}
Detailed expressions of $k_{z,i}^{(\rm II)}$ and $\Phi_i^{(\rm II)}$ are included in Appendix \ref{app:wavefunc}.

    \subsection{Transfer matrix}
    
Wavefunctions in different regions must satisfy the connecting conditions which require the wavefunction and its derivative be continuous at the interfaces between adjacent regions. 

    \subsubsection{The interface between regions I and II}

We first consider the interface between regions I and II where $z$-coordinate is 0 ($z=0$). 
One thing worth to emphasize is that 
since the effective mass is spatially varying, 
the kinetic term along $z$-direction has to be modified as  
\begin{eqnarray}
-\frac{\hbar^2}{2m}\partial_z^2\rightarrow -\frac{\hbar^2}{2}\dif{}{z}\bigg(\frac{1}{m(z)}\dif{}{z}\bigg),
\label{eq:mass_modify}
\end{eqnarray}
in order to ensure hermiticity across the interfaces. 
Adopting Eq. (\ref{eq:mass_modify}) and imposing the continuity conditions for the wavefunction and its derivative, we obtain 
\begin{eqnarray}
\Psi^{(\rm I)}(z=0)&=&\Psi^{(\rm II)}(z=0) \nonumber\\
\frac{1}{m_0}\partd{}{z}\Psi^{(\rm I)}(z=0)&=&\frac1m\partd{}{z}\Psi^{(\rm II)}(z=0).
\label{eq:continuity_I_II}
\end{eqnarray}
   
Plugging the expressions of the wavefunctions in region I and II in Eq. (\ref{wavef}) into the continuity conditions in Eq. (\ref{eq:continuity_I_II}),
we arrive at
\begin{eqnarray}
    	\begin{split}
    	\begin{pmatrix}
    		\Phi^{(\rm I)} & \Phi^{(\rm I)} \\
    		\frac{\ri}{m_0}\Phi^{(\rm I)}\bm{K}_z^{(\rm I)} & -\frac{\ri}{m_0}\Phi^{(\rm I)}\bm{K}_z^{(\rm I)}
    	\end{pmatrix}
    	\begin{pmatrix}
    		A^{(\rm I)}\\
    		B^{(\rm I)}
    	\end{pmatrix}
    	\\=
    	\begin{pmatrix}
    		\Phi^{(\rm II)} & \Phi^{(\rm II)} \\
    		\frac{\ri}{m}\Phi^{(\rm II)}\bm{K}_z^{(\rm II)} & -\frac{\ri}{m}\Phi^{(\rm II)}\bm{K}_z^{(\rm II)}
    	\end{pmatrix}
    	\begin{pmatrix}
    		A^{(\rm II)}\\
    		B^{(\rm II)}
    	\end{pmatrix}
    	\end{split}, 
\end{eqnarray}
in which $A^{(\Gamma)}$ and $B^{(\Gamma)}$ are four-component column vectors defined as
\begin{eqnarray}
A^{(\Gamma)}&=&(A_1^{(\Gamma)},\cdots,A_4^{(\Gamma)})^T,\nonumber\\
B^{(\Gamma)}&=&(B_1^{(\Gamma)},\cdots,B_4^{(\Gamma)})^T,
\end{eqnarray}
and $\Phi^{(\Gamma)}$ and $\bm{K}_z^{(\Gamma)}$ are $4\times 4$ matrices defined as
\begin{eqnarray}
\Phi^{(\Gamma)}&=&(
    \Phi_1^{(\Gamma)}, \cdots, \Phi_4^{(\Gamma)})\nonumber,\nonumber\\
\bm{K}_z^{(\Gamma)}&=&{\rm diag}(
	    k_{z,1}^{(\Gamma)},k_{z,2}^{(\Gamma)},k_{z,3}^{(\Gamma)},k_{z,4}^{(\Gamma)}). 
\end{eqnarray}

For convenience, we introduce $\mathcal{M}_1^{(\Gamma)}$ defined as 
\begin{equation}
    	\mathcal{M}_1^{(\Gamma)}=\begin{pmatrix}
    		\Phi^{(\Gamma)} & \Phi^{(\Gamma)} \\
    		\frac{\ri}{m_{\Gamma}}\Phi^{(\Gamma)}\bm{K}_z^{(\Gamma)} & -\frac{\ri}{m_{\Gamma}}\Phi^{(\Gamma)}\bm{K}_z^{(\Gamma)}
    	\end{pmatrix},
\end{equation}
which  gives
    \begin{equation}
        \begin{pmatrix}
        	A^{(\rm II)} \\
        	B^{(\rm II)}
        \end{pmatrix}=\bm{M}_1\begin{pmatrix}
        A^{(\rm I)}\\
        B^{(\rm I)}
        \end{pmatrix},
    \end{equation} 
where the $8\times 8$ matrix $\bm{M}_1$ is defined as
    \begin{equation}
    	\bm{M}_1=(\mathcal{M}_1^{(\rm II)})^{-1}\mathcal{M}_1^{(\rm I)}. 
    \end{equation}

\subsubsection{The interface between regions II and III}

The treatment for the interface between region II and III where the $z$-coordinate is $d$ ($z=d$) is similar. This gives
    \begin{equation}
    	\begin{pmatrix}
    		A^{(\rm III)} \\
    		B^{(\rm III)}
    	\end{pmatrix}=\bm{M}_2\begin{pmatrix}
    		A^{(\rm II)}\\
    		B^{(\rm II)}
    	\end{pmatrix},
    \end{equation}
in which the $8\times 8$ matrix $\bm{M}_2$ is defined as 
    \begin{equation}
    	\begin{split}
    		\bm{M}_2=(\mathcal{M}_2^{(\rm III)})^{-1}\mathcal{M}_2^{(\rm II)}, 
    	\end{split}
    \end{equation}
where
    \begin{equation}
    	\mathcal{M}_2^{(\Gamma)}=\begin{pmatrix}
    		\Phi^{(\Gamma)}\re^{\ri \bm{K}_z^{(\Gamma)}d} & \Phi^{(\Gamma)}\re^{-\ri \bm{K}_z^{(\Gamma)}d} \\
    		\frac{\ri}{m_\Gamma} \Phi^{(\Gamma)}\bm{K}_z^{(\Gamma)}\re^{\ri \bm{K}_z^{(\Gamma)} d} & -\frac{\ri}{m_\Gamma} \Phi^{(\Gamma)}\bm{K}_z^{(\Gamma)}\re^{-\ri \bm{K}_z^{(\Gamma)}d}
    	\end{pmatrix}.
    \end{equation}

\subsubsection{Transfer matrix connecting regions I and III}
   
    The relation between the coefficients in region I and region III  can then be obtained through matrix multiplication,
    \begin{equation}
    	\begin{pmatrix}
    		A^{(\rm III)}\\
    		B^{(\rm III)}
    	\end{pmatrix}=\bm{M}\begin{pmatrix}
    	A^{(\rm I)}\\
    	B^{(\rm I)}
    	\end{pmatrix},
    \end{equation}
    where the $8\times 8$ matrix $\bm{M}$ is defined as
\begin{equation}
    \bm{M}=\bm{M}_2\bm{M}_1.
\end{equation}

\subsection{Transmission coefficients}

If the waves are incident from region I, 
then there is only outgoing but no incoming wave in region III.
Accordingly, $B^{(\rm III)}$ has to be set as zero. 
From the following equation of transfer matrix 
\begin{equation}
    	\begin{pmatrix}
    		A^{(\rm III)}\\
    		0
    	\end{pmatrix}
    	=\bm{M}\begin{pmatrix}
    		A^{(\rm I)}\\
    		B^{(\rm I)}
    	\end{pmatrix},
\end{equation}
we obtain
    \begin{equation}
    	\begin{pmatrix}
    	    -M_{AB} & \mathbf{1} \\
    	    -M_{BB} & 0
    	\end{pmatrix}\begin{pmatrix}
    	B^{(\rm I)} \\
    	A^{(\rm III)}
    	\end{pmatrix}=\begin{pmatrix}
    	    M_{AA}A^{(\rm I)}\\
    	    M_{BA}A^{(\rm I)}
    	\end{pmatrix},
	\label{eq:rearragne_transfer_Mat}
    \end{equation}
where $M_{AA},M_{AB},M_{BA},M_{BB}$ are $4\times 4$ matrices defined according to
\begin{equation}
\bm{M}=\begin{pmatrix}
    	M_{AA} & M_{AB}\\
    	M_{BA} & M_{BB}\\
    	\end{pmatrix}.
\end{equation}
From Eq. (\ref{eq:rearragne_transfer_Mat}), $B^{(\rm I)}$ and $A^{(\rm III)}$  can be solved  as 
\begin{eqnarray}
    	B^{(\rm I)}&=&-M_{BB}^{-1}M_{BA}A^{(\rm I)}\nonumber \\
    	A^{(\rm III)}&=&(M_{AA}-M_{AB}M_{BB}^{-1}M_{BA})A^{(\rm I)}.
    	\label{coef}
\end{eqnarray}
In this way,  the transmitted and reflected wavefunctions (corresponding to coefficients $B^{(\rm I)}$ and $A^{(\rm III)}$) can be fully determined from the incident wavefunction (corresponding to coefficients $A^{(\rm I)}$).


Probability current $J_z$ along $z$-direction is defined as
\begin{eqnarray}
J_z=-\frac{\ri \hbar}{2m}(\Psi^\dagger\partd{\Psi}{z}-\partd{\Psi^\dagger}{z}\Psi), 
\end{eqnarray}
where the wavefunction $\Psi(z)$ is given in Eq. (\ref{wavef}). 
The probability current  $J_z^{(\text{I})}$ in region I can be expressed as
\begin{eqnarray}
J_z^{(\text{I})}&=&J_z^{\rm in}-J_z^{\rm refl}\nonumber\\
&=&\sum_{i=1}^4 \frac{k_{z,i}^{(\text{I})}}{m_0}(|A^{(\text{I})}_i|^2-|B^{(\text{I})}_i|^2)[\Phi_i^{(\text{I})}]^\dagger\Phi_i^{(\text{I})},
\end{eqnarray}
and the probability current  $J_z^{(\text{III})}$ in region III is
\begin{eqnarray}
J_z^{(\text{III})}=J_z^{\rm trans}
=\sum_{i=1}^4 \frac{k_{z,i}^{(\text{III})}}{m_0}|A^{(\text{III})}_i|^2[\Phi_i^{(\text{III})}]^\dagger\Phi_i^{(\text{III})}.
\end{eqnarray}
The reflection and transmission probabilities are defined as the ratios of the reflected and transmitted probability currents to the incident current, i.e. 
\begin{eqnarray}
R=\frac{J_z^{\rm refl}}{J_z^{\rm in}},\,\,\,T=\frac{J_z^{\rm trans}}{J_z^{\rm in}},
\end{eqnarray}
which satisfies   current conservation $R+T=1$ for propagating modes.

\section{Magnetoresistance and spin polarization}
	\label{sec:magneto}
	
\subsection{Conductance and magnetoresistance}

Having obtained the spin-dependent transmission and reflection coefficients in the previous section, we now turn to the calculation of the conductance of the spin valve. 
We restrict to the situation of parallel/anti-parallel magnetizations in the two electrodes.
In such cases, the ${\rm K}$ and ${\rm K}^\prime$ valleys contribute equally as discussed in Sec. \ref{sec:ham}.

In a realistic vertical spin valve device, the experimentally measured total area resistance, $AR$, consists of two distinct contributions: the bulk resistance of the ferromagnetic electrodes and the interface resistance associated with the spacer region. We emphasize that treating the central WSe$_2$ layer effectively as an interfacial scattering region is valid under the condition that its thickness is much smaller than the phase coherence length and the spin diffusion length. Under this assumption, the macroscopic total area resistance can be modeled as a classical series circuit:
\begin{equation}
	AR=\rho_{\rm I} d_{\rm I}+AR_{\rm II}+\rho_{\rm III}d_{\rm III}
	\label{eq:totalAR}
\end{equation}
where $\rho_{\rm I}$ and $\rho_{\rm III}$ denote the bulk resistivities of the electrodes in regions I and III, $d_{\rm I}$ and $d_{\rm III}$ are their respective transport lengths, and $AR_i$ represents the effective interface resistance arising from spin-dependent scattering.

To evaluate the interface resistance, we start from the Landauer formalism in the ballistic transport limit, which assumes that phase coherence is preserved across the thin spacer. 
In this case, the  interface conductance is given by:
	\begin{equation}
		G_{\rm II}=\frac{2e^2}{h}\sum\limits_{i,\bm{k}_\parallel}T_i(E_f,\bm{k}_{\parallel}),
		\label{eq:LB}
	\end{equation}
in which $T_i(E,\bm{k}_\parallel)$ denotes the transmission coefficient for an electron with an in-plane momentum $\bm{k}_\parallel=(k_x,k_y)^T$ and four-component spinor index $i$, 
and the factor of 2 arises from the equal contributions of the two valleys.
In the continuum limit, the summation over transverse modes is replaced by an integral over the two-dimensional Brillouin zone:
\begin{equation}
	G_{\rm II}=\frac{2e^2}{h}\frac{A}{(2\pi)^2}\sum\limits_i\iint{\rm d}^2\bm{k}_\parallel T_i^\sigma(E_f,\bm{k}_\parallel).
	\label{eq:LB_int}
\end{equation}
The integration is performed over a neighborhood of the relevant valley points. 
For practical calculations, a finite momentum cutoff $k_c$ is imposed to restrict the integration to the region where the low-energy effective Hamiltonian remains valid.

Crucially, the resistance derived directly from the Landauer conductance ($R_{\rm L} = 1/G_{\rm II}$) includes not only the intrinsic scattering resistance of the interface but also the Sharvin contact resistance associated with the finite number of transverse propagating modes in the ideal semi-infinite leads. 
To incorporate this consistently into the macroscopic series resistor model (i.e., Eq. (\ref{eq:totalAR})), the Sharvin contribution should be explicitly subtracted \cite{schep1997interface,bauer2002scattering}. 
Therefore, 
the interface area resistance that should be plugged in Eq. (\ref{eq:totalAR}) is described by the following Schep-Bauer formula,
\begin{equation}
	AR_{\rm II}=\frac{Ah}{2e^2}\left(\frac{1}{\sum T}-\frac{1}{N}\right),
	\label{eq:Schep}
\end{equation}
where $N$ is the number of conduction channels in the ferromagnetic electrodes, and $\sum T$ represents the total transmission probability integrated over $\bm{k}_\parallel$ with cutoff $k_c$ in one of the two valleys.

For simplicity, in the following we use the resistance defined directly from the Landauer conductance \(R_{\rm L}=1/G_{\rm II}\) as an effective junction resistance where $G_{\rm II}$ is given in Eq.~(\ref{eq:LB_int}). 
This quantity retains the contribution from spin-dependent coherent scattering across the WSe$_2$ spacer and the two electrode/spacer interfaces, and it also contains the Sharvin contact resistance associated with the ideal leads. On the other hand, it does not include the bulk diffusive resistance of the ferromagnetic electrodes. 
Accordingly, the magnetoresistance presented below characterizes the coherent transport contribution of the junction region, rather than the full device resistance measured in experiment.
However, for the purpose of discussing the negative magnetoresistance at a qualitative level, this approximation is sufficient, since the bulk diffusive resistance of the electrodes is not expected to generate the feature of negative magnetoresistance and mainly provides a spin-independent background contribution.

To evaluate the magnetoresistance of the vertical WSe$_2$-based spin valve, we consider two distinct magnetic configurations corresponding to the parallel (P) and antiparallel (AP) alignments of the ferromagnetic electrodes. In the P configuration, the exchange splitting fields in both electrodes are aligned in the same direction (e.g., $h_{\rm I} = h_{1}$, $h_2=0$). Conversely, in the AP configuration, the relative orientations of the exchange fields are reversed (e.g., $h_{\rm I} = -h_{1}$, $h_2=0$). The total interface conductances for the P and AP states, denoted as $G_P$ and $G_{AP}$ respectively, can be straightforwardly calculated using Eq.~(\ref{coef}).
The magnetoresistance ratio is defined according to the standard formula:
\begin{equation}
	{\rm MR}=\frac{R_{AP}-R_{P}}{R_{P}}=\frac{G_{P}-G_{AP}}{G_{AP}}
	\label{eq:mrform}
\end{equation}
This definition enables a quantitative evaluation of the magnetoresistance  as a function of the relative magnetization alignment.
The above framework provides a basis for the subsequent numerical analysis.

\subsection{Density of states and spin polarization}
\label{subsec:detail_fitting}

In this section, we describe in detail the determination of $k_c$ and $\epsilon_F$ following the method described in Sec. \ref{subsec:model_parameters}. 
Since we will focus on the parallel/anti-parallel cases (i.e., $h_2=0$ in Eq. (\ref{HamIII})),
only the ${\rm K}$ valley is retained.  
The inclusion of ${\rm K}^\prime$ valley  just contributes a factor of $2$. 

In the region of bottom electrode  (e.g., in region I in Fig. \ref{fig:verticalspinvalve}), the spin-resolved density of states (DOS) can be obtained by
	\begin{equation}
		D_i(E)=\frac{1}{(2\pi)^3}\int \di^3\vec{k}\delta(E-E_i(\vec{k})), 
		\label{eq:DOS}
	\end{equation}
in which $E_i(\vec{k})$ is given by
\begin{equation}
		E_i(\vec{k})=\mathcal{E}_i^{\rm (I)}+\frac{\hbar^2k_{z,i}^{(\rm I)2}}{2m_0}-\varepsilon_F.
\end{equation}
Since $S_x$ is a good quantum number in region I,
the index $i=1,2,3,4$ in Eq. (\ref{eq:DOS}) can be chosen to label the four spin-split bands in region I.
The convention is that bands $i=1,3$ possess right spin corresponding to eigenvalue $S_x=1$, while bands $i=2,4$ possess right spin corresponding to eigenvalue $S_x=-1$.
Denote  $D_\leftarrow=D_2+D_4$ and  $D_\rightarrow=D_1+D_3$
as spin left and spin right DOS, respectively. 
Then the spin polarization in the electrodes can be determined as
\begin{equation}
P=\frac{D_\rightarrow-D_\leftarrow}{D_\rightarrow+D_\leftarrow}.
\label{eq:expression_spin_polarization}
\end{equation}
Detailed calculations of DOS are included in Appendix \ref{app:DOS}.

As discussed in Sec. \ref{subsec:model_parameters}, spin polarization and magnetoresistance are both experimentally measurable quantities. 
Hence combining Eq. (\ref{eq:expression_spin_polarization}) with Eq. (\ref{eq:mrform}),
${\rm MR}(k_c,\varepsilon_F)$ and $P(k_c,\varepsilon_F)$ can both be expressed as functions of the momentum space cutoff $k_c$ and Fermi energy $\epsilon_F$. 
Using the experimental values of magnetoresistance (approximately -0.8\%, monolayer) and spin polarization (approximately 16.7\%)  reported in Ref. \onlinecite{wang2022spin}, 
and performing a systematic numerical scan over physically reasonable ranges of $k_c$ and $\varepsilon_F$,
we obtain
$k_c=0.105 \text{\AA}^{-1}$ and $\varepsilon_F=0.535 {\rm eV}$.
In the numerical scan, the exchange field $h_1=\pm h_I$ in regions I and III and the gate potential $V_g$ are taken as $h_I=0.25$eV and $V_g=0.57$eV as discussed in Sec. \ref{sec:numer}.

	\section{Numerical results}
	\label{sec:numer}
    In this section, we present the numerical results of spin-dependent transport in the graphene/WSe$_2$/graphene vertical structure based on the aforementioned  framework.
    
\subsection{Dependence of magnetoresistance  on layer number}
\label{subsec:dependence_layer_num}

\begin{figure}
    	\centering
    	\includegraphics[width=1\linewidth]{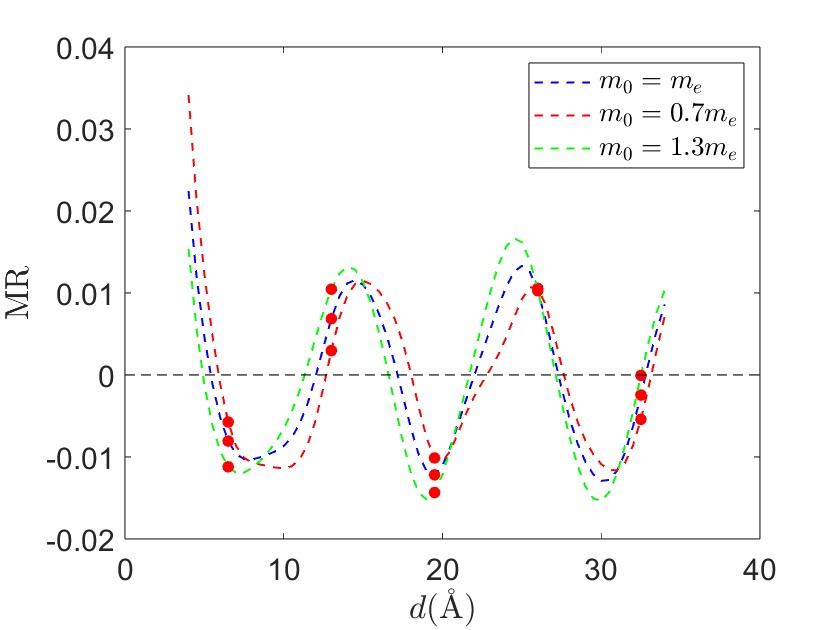}
    	\caption{Magnetoresistance (MR) as a function of the \Ws thickness \(d\). The blue, red, and green curves correspond to calculations with \(d\) treated as a continuous variable for \(m_0 = m_e\), \(0.7m_e\), and \(1.3m_e\), respectively. The red dots represent the MR values for discrete layer numbers, namely for \(d = n d_0\) with integer \(n\) and \(d_0 = 6.5\,\text{\AA}\).}
    	\label{fig:mrd}
    \end{figure}

In experiments, the thickness $d$ of the spacer region of WSe$_2$ between the electrodes can be tuned by changing the number of WSe$_2$ layers. 
We will consider $d$ as a continuous variable in our calculations, 
though in practice, $d$ is equal to $Nd_0$, where $N$ is the layer number and $d_0$ is the thickness of monolayer WSe$_2$.
In numerical calculations, the exchange field $h_I$ in region I is chosen as a moderate value of $h_{\rm I}=0.25 {\rm eV}$;
the exchange field in region III is chosen as $(h_1=\pm h_I,h_2=0)$
where ``$+$" and ``$-$"  correspond to the parallel and anti-parallel configurations, respectively;
the gate potential is taken as $V_g=0.57{\rm eV}$, so that the Fermi level in region II is located on the in-plane valence band maximum;
and the effective mass in \Ws is chosen as $m=1.6m_e$, where $m_e$ is free electron mass\cite{wang2022spin}.
As for the effective mass $m$ in regions I and III,
we have tuned its value from $0.6m_e$ to $1.4m_e$,
finding that the magnetoresistance is only weakly dependent on $m$ in this range as shown in Fig. \ref{fig:mrd} for several representative values of $m$.
Hence, $m_e$ is adopted as the effective mass for regions I and III in all subsequent calculations. 
    
Fig. \ref{fig:mrd} shows the magnetoresistance MR in Eq. (\ref{eq:mrform}) as a function of WSe$_2$ thickness $d$,
exhibiting  an oscillatory behavior, with its sign alternating between positive and negative values as thickness increases. 
The oscillation behavior is consistent with reported experimental results in Ref. \onlinecite{wang2022spin}.

The oscillating behavior of magnetoresistance with respect to inter-layer thickness $d$ 
can be clearly understood from a semiclassical picture in the limit of $d\gg \lambda_F$,
where $\lambda_F$ is the Fermi wavelength. 
As an electron incident from region I traverses region II in Fig. \ref{fig:verticalspinvalve}, 
its spin precesses around the effective magnetic field arising from spin-orbit coupling in the WSe$_2$ material, accumulating a precession angle proportional to thickness $d$. 
Upon reaching the interface between region II and III, the probability of transmission into the top electrode (i.e., region III) is determined by the degree of relative alignment between the spin direction of the electron and the exchange field in region III,
having the largest and smallest transmission coefficients for parallel and anti-parallel alignments, respectively. 
In particular, when the spins of the outgoing electrons are flipped by the effective Zeeman field in the inter-layer WSe$_2$ material for special $d$'s,
the electron spins match and mis-match with the directions of exchange fields in the anti-parallel and parallel configurations, respectively,
resulting in negative magnetoresistance according to Eq. (\ref{eq:mrform}).

On the other hand, there is a quantum mechanical mechanism for negative magnetoresistance arising from interference effects, not of a semiclassical origin, which will be discussed in details in Sec. \ref{subsec:interference_MR}.
In particular, the Fabry-Pérot-like interference contribution to the magnetoresistance can dominate over semiclassical considerations in certain cases, 
such that the system exhibits negative magnetoresistance even if the magnetoresistance is expected to be positive from  a semiclassical estimation.  


\subsection{Dependence of magnetoresistance on gate potential}

\begin{figure}
    	\centering
    	\includegraphics[width=1\linewidth]{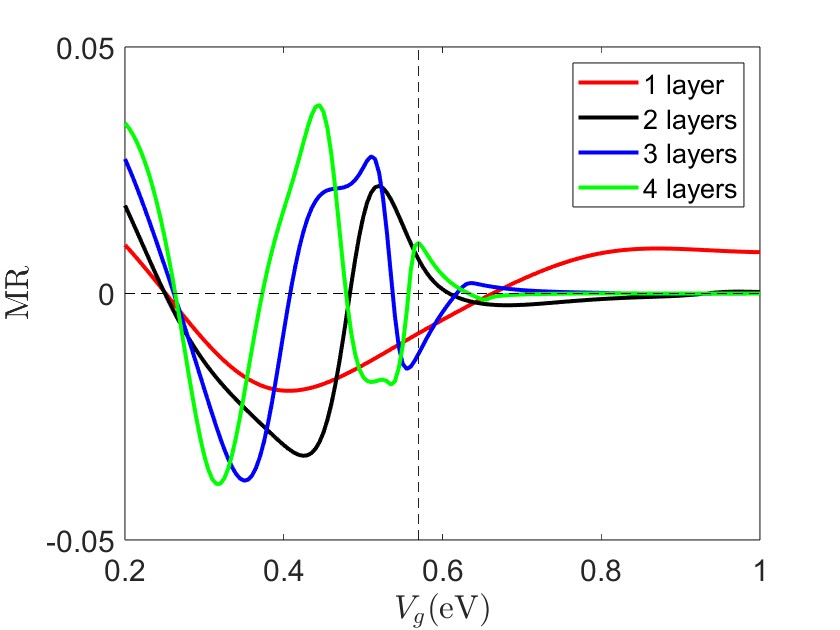}
    	\caption{Magnetoresistance MR as a function of $V_g$ for monolayer (red), bilayer (black), trilayer (blue) and four-layer (green) WSe$_2$. The vertical dashed line indicates the value of the gate potential $V_g=0.57{\rm eV}$.}
    	\label{fig:mrvg}
    \end{figure}

Next, we investigate the dependence of magnetoresistance on the gate potential $V_g$. 
We have calculated the $V_g$-dependence for 1 to 4 layers of WSe$_2$,
keeping all other parameters the same as Sec. \ref{subsec:dependence_layer_num}.
As shown in Fig. \ref{fig:mrvg}, when $V_g$ is tuned to values near the valence band maximum of \Ws (around the vertical dashed line), the magnetoresistance alternates between positive and negative values as the layer number increases, which is consistent with the thickness-dependent oscillatory behavior discussed in Sec. \ref{subsec:dependence_layer_num}. 
As $V_g$ moves away from the in-plane valence band maximum, the magnetoresistance gradually approaches a positive value and the oscillatory behavior by changing layer number becomes suppressed or even vanishes.
 
    \par 
    \subsection{Dependence of magnetoresistance  on exchange field}
    
\begin{figure}
    	\centering
    	\includegraphics[width=1\linewidth]{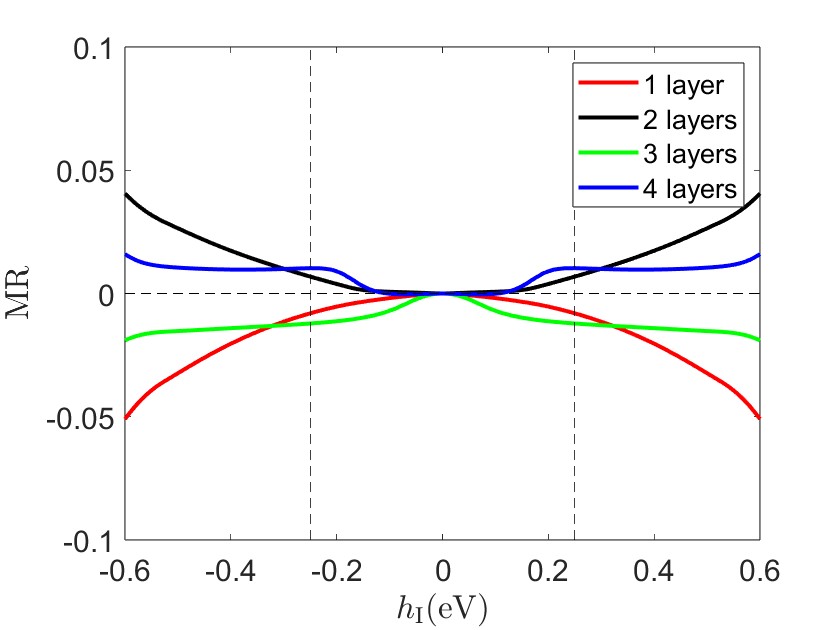}
    	\caption{Magnetoresistance MR as a function of exchange field $h_{\rm I}$ for monolayer (red), bilayer (black), trilayer (blue) and four-layer (green)\Ws. The vertical dashed linea indicate the value $h_{\rm I}=\pm 0.25{\rm eV}$.}
    	\label{fig:mrh}
    \end{figure}

We also study the dependence of the magnetoresistance on the exchange field. 
The exchange field is applied in regions I and III, with field value $h_I$ along $x$-direction in region I and $(h_1=\pm h_{\rm I},h_2=0)$ in region III.
Fig. \ref{fig:mrh} shows magnetoresistance as a function of $h_{\rm I}$ for several layer numbers,
in which the field $h_{\rm I}$ varies from $-0.6$eV to $0.6$eV.
The dependence  of magnetoresistance on $h_{\rm I}$ has an approximately parabolic behavior when the exchange field is much smaller than the Fermi energy, i.e., $|h_{\rm I}|\ll\varepsilon_F$.

\section{Fabry-Pérot  interference contribution to magnetoresistance}
\label{subsec:interference_MR}

In the absence of SOC in the inter-layer material (i.e., in region II),
there is no effective  Zeeman field in the spacer region. 
As a result,
one would expect from the semiclassical picture that the tunneling conductance in the parallel configuration always exceeds that in the anti-parallel  configuration. 
However, we will show that because of interference effects between the two interfaces,
the above intuitive expectation does not always hold, 
and transmission in anti-parallel configuration can be larger than parallel configuration for certain thicknesses. 
This counterintuitive $G_{AP}>G_P$ behavior demonstrates that coherent multiple reflections in Fabry-Pérot-like interference alone can lead to a sign reversal of magnetoresistance at specific thicknesses.
    
To isolate the role of coherent multiple reflections from the SOC-induced spin precession, 
a simplified spin-conserving model is considered in which SOC is turned off ($\lambda=0$) and only normal incidence at the K point is retained (i.e., $k_x=k_y=0$). 
Effective masses  are taken as $m=m_0=m_e$ for simplification, which does not affect the Fabry-Pérot interference physics.
In this case, $\sigma_z$ is a good quantum number and the tunneling problem can be decomposed into mutually independent $\sigma_z$-resolved scattering channels,
where $\sigma_z$ represents the orbital degrees of freedom as discussed below Eq. (\ref{HamI}).
When $\sigma_z=1$, the energy is around the conductance band minimum, lying far above the Fermi energy.
As a result, the transmission is small.
On the other hand,  when $\sigma_z=-1$, the energy is around the valence band maximum where the Fermi energy locates,
having large transmission coefficients. 
We will focus on the $\sigma_z=-1$ with large transmssions in what follows. 

When $\sigma_z=-1$, the Hamiltonians in different regions are
    \begin{eqnarray}
    		H_{\rm I}&=&-h_{\rm I}S_x-\frac{\hbar^2}{2m}\partial_z^2\nonumber\\
    		H_{\rm II}&=&-\frac{\hbar^2}{2m}\partial_z^2+U\nonumber\\
    		H_{\rm III}&=&\pm h_{\rm I}S_x-\frac{\hbar^2}{2m}\partial_z^2,
		\label{eq:Ham_FP}
    \end{eqnarray}
in which "$-$" is for P state and "+" is for AP state,
and $U$ is given by
\begin{eqnarray}
U=-\frac{\Delta}{2}+V_g+\varepsilon_F.
\end{eqnarray}
The wavefunction takes the  form
\begin{equation}
    	\Psi(z)=\sum\limits_{s_z=\pm 1}\psi_{s_z}(z)\chi_{s_z}, 
\end{equation}	
    where $\psi_{s_z}(z)$ and $\chi_{s_z}$ are the spatial and spin parts of the wavefunction, respectively. 

\begin{figure}
\centering
\includegraphics[width=8.5cm]{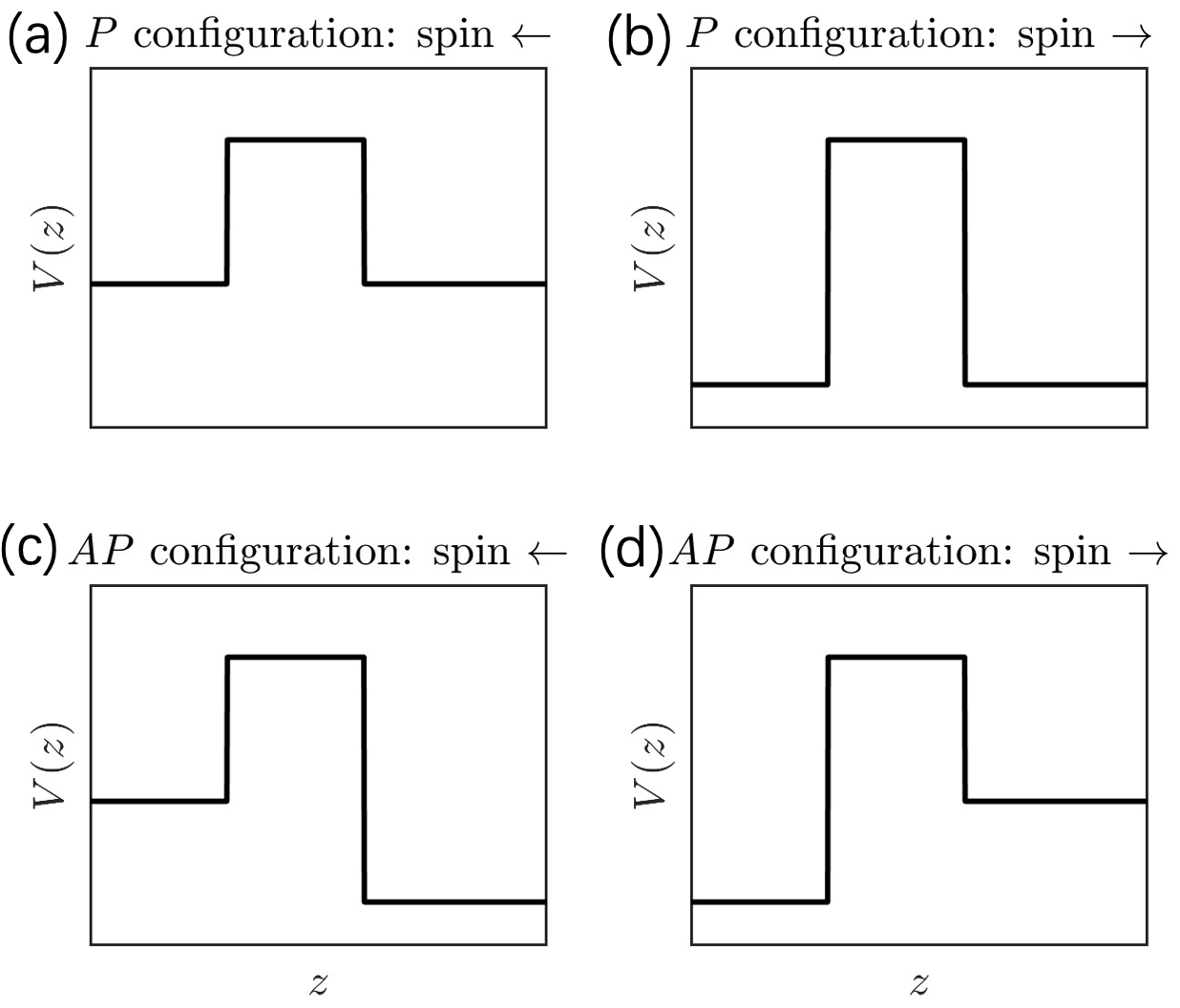}
\caption{Spin-resolved potential files
for (a) spin left in P configuration, (b) spin right in P configuration,
(c) spin left in AP configuration, (d) spin right in AP configuration.}
\label{fig:spin_potential}
\end{figure}

    	\begin{figure}
    	\centering
    	\includegraphics[width=8.5cm]{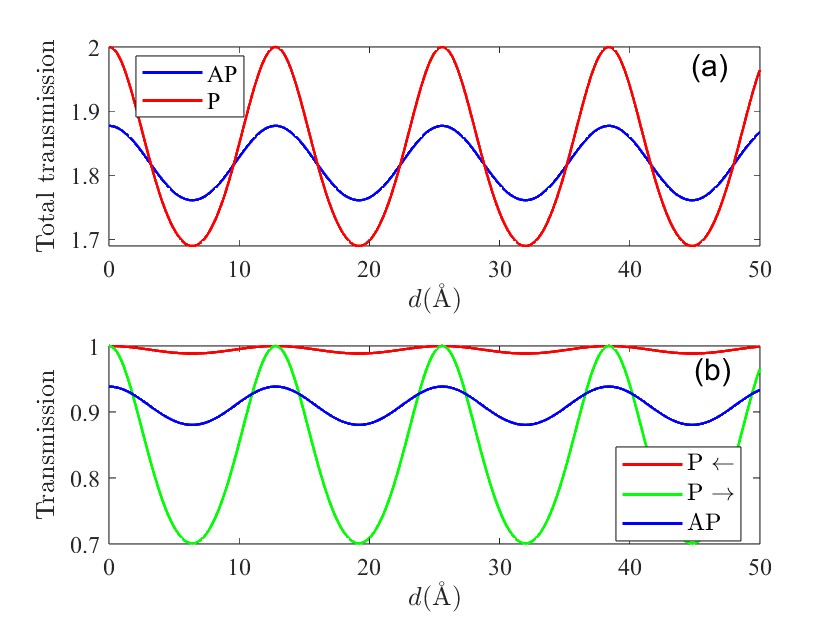}
    	\caption{Spin-dependent transmission in the simplified model. 
(a) total transmission for P and AP alignments,
and (b) spin-resolved transmissions $T_P^\leftarrow,T_P^\rightarrow,T_{AP}^{\leftarrow},T_{AP}^{\rightarrow}$.}
    	\label{fig:transtotal}
    \end{figure}

As shown in Fig. \ref{fig:spin_potential},
in the parallel configuration, 
regions I and III have the same potential for both spin channels, forming symmetric structures of square potential;
whereas in the anti-parallel configuration, 
there is an imbalance between the potentials in regions I and III, because of the sign difference in $S_x$ terms in Eq. (\ref{eq:Ham_FP}).
This difference lies at the heart of the distinction of the interference  behaviors between parallel and anti-parallel configurations.

The transmission probability can be analyzed in terms of multiple reflections and interference, analogous to a Fabry–Pérot resonator. 
In the parallel configuration, for a given spin channel $\lambda$ ($\lambda=\leftarrow,\rightarrow$), the top and bottom interfaces act as partial reflectors, with reflection coefficients 
\begin{eqnarray}
    	r^\lambda_{12}&=&\frac{k_\lambda-k_{\rm II}}{k_\lambda+k_{\rm II}}\nonumber\\
	r^\lambda_{21}&=&-r^\lambda_{12}\nonumber\\
	r^\lambda_{23}&=&\frac{k_{\rm II}-k_\lambda}{k_{\rm II}+k_\lambda},
\end{eqnarray}
    and transmission coefficients
\begin{eqnarray}
    	t^\lambda_{12}&=&\frac{2k_\lambda}{k_\lambda+k_{\rm II}}\nonumber\\
	t^\lambda_{23}&=&\frac{2k_{\rm II}}{k_{\rm II}+k_\lambda},
\end{eqnarray}
in which 
\begin{eqnarray}
k_\leftarrow&=&\sqrt{\frac{2m}{\hbar^2}(E-h_{\rm I})}\nonumber\\ 
k_\rightarrow&=&\sqrt{\frac{2m}{\hbar^2}(E+h_{\rm I})}\nonumber\\ 
k_{\rm II}&=&\sqrt{\frac{2m}{\hbar^2}(E-U)}.
\label{eq:expressions_k_FP}
\end{eqnarray}

The transmission amplitude through the barrier can be obtained by summing over all possible paths that involve $n$ round trips inside the barrier,
\begin{eqnarray}
    	t^\lambda&=&t^\lambda_{12}t^\lambda_{23}\re^{\ri k_{\rm II}d}\sum\limits_{n=0}^\infty(r^\lambda_{23}r^\lambda_{21}\re^{2\ri k_{\rm II }d})^n\nonumber\\
	&=&\frac{t^\lambda_{12}t^\lambda_{23}\re^{\ri k_{\rm II}d}}{1-(r^\lambda_{12})^2\re^{2\ri k_{\rm II}d}}.
\end{eqnarray} 
Transmission probabilities $T^\lambda_P$ can be obtained by taking square of $t^\lambda$:
\begin{eqnarray}
T_P^\lambda&=&\frac{4k_\lambda^2k_{\rm II}^2}{4k_\lambda^2k_{\rm II}^2+(k_{\rm II}^2-k_{\lambda}^2)^2\sin^2(k_{\rm II}d)}.
\label{transp}
\end{eqnarray}
When $k_{\rm II}d=m\pi$, transmission probabilities reach unity $T_P^\lambda=1$ ($\lambda=\leftarrow,\rightarrow$),
which is a consequence of constructive interference, corresponding to perfect transmission;
whereas when $k_{\rm II}d=(m+\frac{1}{2})\pi$, 
transmission probabilities reach the smallest values $T_P^\lambda=\frac{4k_\lambda^2k_{\rm II}^2}{(k_{\rm II}^2+k_{\lambda}^2)^2}$ because of destructive interference.

In the anti-parallel configuration, 
the reflection and transmission coefficients at the two interfaces are given by
\begin{eqnarray}
r^\lambda_{12}&=&\frac{k_\lambda-k_{\rm II}}{k_{\rm II}+k_{\lambda}}\nonumber\\
r^\lambda_{21}&=&-r^\lambda_{12}\nonumber\\
r^\lambda_{23}&=&\frac{k_{\rm II}-k_{-\lambda}}{k_{\rm II}+k_{-\lambda}}.
\end{eqnarray}
and 
\begin{eqnarray}
t^\lambda_{12}&=&\frac{2k_\lambda}{k_\lambda+k_{\rm II}}\nonumber\\
t^\lambda_{23}&=&\frac{2k_{\rm II}}{k_{\rm II}+k_{-\lambda}}.
\end{eqnarray}
Notice that the asymmetric appearances of $k_{\lambda}$ in the formulas at interface ``$12$"
and $k_{-\lambda}$ at interface ``$23$"
are the consequence of the asymmetric potentials in Fig. \ref{fig:spin_potential} (c,d) for the anti-parallel configuration.

Similar calculations of coherent sum over multiple reflection and transmission contributions show that the transmission probabilities in the AP case are given by
\begin{flalign}
&T_{AP}^{\leftarrow}=T_{AP}^{\rightarrow}=\nonumber\\
&\frac{4k_\leftarrow k_\rightarrow k_{\rm II}^2}{k_{\rm II}^2(k_\leftarrow+k_\rightarrow)^2\cos^2(k_{\rm II }d)+(k_{\rm II}^2+k_\leftarrow k_\rightarrow)^2\sin^2(k_{\rm II}d)}.
\label{eq:T_AP_lr_arrows}
\end{flalign}
Notice that compared with the parallel configuration,
when $k_{\rm II}d=m\pi$, the constructive interference gives a smaller value less than unity,
$T_{AP}^{\leftarrow}=T_{AP}^{\rightarrow}=\frac{4k_\leftarrow k_\rightarrow}{(k_\leftarrow+k_\rightarrow)^2}$.
On the other hand, when $k_{\rm II}d=(m+\frac{1}{2})\pi$, the destructive interference gives a larger value 
$T_{AP}^{\leftarrow}+T_{AP}^{\rightarrow}=\frac{8k_\leftarrow k_\rightarrow k_{\rm II}^2}{(k_{\rm II}^2+k_\leftarrow k_\rightarrow)^2}\geq T_{P}^{\leftarrow}+T_{P}^{\rightarrow}=\sum_{\lambda}\frac{4k_\lambda^2k_{\rm II}^2}{(k_{\rm II}^2+k_{\lambda}^2)^2}$
provided that $k_{\leftarrow}k_{\rightarrow}<(2+\sqrt{3})k_{\rm II}^2$
 (for a proof of this inequality, see Appendix \ref{app:proof_inequal}). 
Intuitively, the reason for the above phenomena is that the asymmetric structure of the potential in the anti-parallel  configuration suppresses not only the constructive but also the destructive interference effects.
As a result, for values of thickness around the destructive interference conditions $k_{\rm II}d=(m+\frac{1}{2})\pi$,
it is possible for the transmission probability in the anti-parallel configuration to be larger than that in the parallel configuration,
leading to negative magnetoresistance.
Notice that this is a pure quantum mechanical phenomenon,
which is beyond the semiclassical picture.

We have checked the interference induced negative magnetoresistance numerically as shown in Fig. \ref{fig:transtotal}.
Fig. \ref{fig:transtotal} (a) plots the transmission probabilities of parallel (line in red) and anti-parallel (line in blue) configurations, respectively;
and Fig. \ref{fig:transtotal} (b) shows the spin-resolved transmission probabilities 
where there is a single line for the anti-parallel configuration, since spin left and spin right have identical tunneling values in the anti-parallel case as can be seen from Eq. (\ref{eq:T_AP_lr_arrows}).
It is clear from Fig. \ref{fig:transtotal} (a) that there are narrow regions where the blue line is above the red line,
which are the parameter regions where magnetoresistance becomes negative.

\section{Summary}
\label{sec:summary}

In summary, we have theoretically investigated spin-dependent transport in a vertical spin valve composed of ferromagnetic doped graphene electrodes and a few-layer WSe$_2$ spacer. Using effective $\bm{k}\cdot\bm{p}$ Hamiltonians and a transfer-matrix scattering approach, we calculated the conductance and magnetoresistance within the Landauer framework. The results reveal clear magnetoresistance oscillations as a function of WSe$_2$ thickness, most pronounced when $V_g$ is near the valence-band maximum. 
In addition, we showed that Fabry-P\'erot-like interference arising from coherent multiple reflections across the spacer can also make an important contribution to the magnetoresistance and, in certain thickness regimes, can lead to negative magnetoresistance in a way that is not captured by the semiclassical picture.
These results provide a theoretical understanding of the behavior of negative magnetoresistance in WSe$_2$-based vertical spin valves and may offer useful insights for the design of tunable spintronic devices.

\begin{acknowledgments}

Y.W. and W.Y. are supported by the Fundamental Research Funds for the Central Universities.
X.W. is supported by the National Key R\&D Program of China (No. 2022YFA1405900).

\end{acknowledgments}

	
	\appendix
	
	\begin{widetext}

\section{Detailed derivation of eigenvalues and eigen-wavefunctions in region II}
\label{app:wavefunc}
		Notice that $[S_z,H_{\rm II}^\prime]=0$, so $S_z$ is a good quantum number. $S_z$ can be block-diagonalized:
	\begin{equation}
		H_{\rm II}^\prime(S_z=\pm 1)=v(k_x\sigma_x+k_y\sigma_y)+\frac12(\Delta\mp\lambda)\sigma_z\pm \frac12 \lambda
	\end{equation}
	Let $M$ be defined as $M=a\sigma_z+b\sigma_x+c\sigma_y$:
	\begin{equation}
		M=\begin{pmatrix}
		    a & b-\ri c\\
		    b+\ri c & -a
		\end{pmatrix}
	\end{equation}
	$\sqrt{a^2+b^2+c^2}$ and $-\sqrt{a^2+b^2+c^2}$ are two eigenvalues of matrix $M$. Then the two eigenvalues of $M$ are
	\begin{equation}
		\begin{split}
		\frac{1}{\sqrt{2\sqrt{a^2+b^2+c^2}(\sqrt{a^2+b^2+c^2}-a)}}\begin{pmatrix}
			b-\ri c\\
			\sqrt{a^2+b^2+c^2}-a
		\end{pmatrix}\\
		\frac{1}{\sqrt{2\sqrt{a^2+b^2+c^2}(\sqrt{a^2+b^2+c^2}+a)}}\begin{pmatrix}
				b-\ri c\\
				-\sqrt{a^2+b^2+c^2}-a
			\end{pmatrix}\\
		\end{split}
	\end{equation}
	Let $a=\frac12(\Delta\mp\lambda),b=vk_x,c=vk_y$, $H_{\rm II}^\prime$ can be expressed as \begin{equation}
		H_{\rm II}^{\prime}=M\pm \frac{1}{2}\lambda
	\end{equation} then we have eigenvalues of $H_{\rm II}^\prime$ \begin{equation}
		\begin{split}
			\mathcal{E}_1^{(\rm II)}=\sqrt{v^2k_{\parallel}^2+\frac14(\Delta-\lambda)^2}+\frac12\lambda \\
			\mathcal{E}_2^{(\rm II)}=\sqrt{v^2k_{\parallel}^2+\frac14(\Delta+\lambda)^2}-\frac12\lambda \\
			\mathcal{E}_3^{(\rm II)}=-\sqrt{v^2k_{\parallel}^2+\frac14(\Delta-\lambda)^2}+\frac12\lambda \\
			\mathcal{E}_4^{(\rm II)}=-\sqrt{v^2k_{\parallel}^2+\frac14(\Delta+\lambda)^2}-\frac12\lambda \\
		\end{split}
		\end{equation}
	and the $z$-momenta in region II are
	\begin{equation}
		\begin{split}
			k_{z,1}^{(\rm II)}=\sqrt{\frac{2m}{\hbar^2}(E-V_g-\sqrt{v^2k_{\parallel}^2+\frac14(\Delta-\lambda)^2}-\frac12\lambda)} \\
			k_{z,2}^{(\rm II)}=\sqrt{\frac{2m}{\hbar^2}(E-V_g-\sqrt{v^2k_{\parallel}^2+\frac14(\Delta+\lambda)^2}+\frac12\lambda)} \\
			k_{z,3}^{(\rm II)}=\sqrt{\frac{2m}{\hbar^2}(E-V_g+\sqrt{v^2k_{\parallel}^2+\frac14(\Delta-\lambda)^2}-\frac12\lambda)} \\
			k_{z,4}^{(\rm II)}=\sqrt{\frac{2m}{\hbar^2}(E-V_g+\sqrt{v^2k_{\parallel}^2+\frac14(\Delta+\lambda)^2}+\frac12\lambda)} \\
		\end{split}
	\end{equation}
	The eigen-wavefunctions can also be obtained from the eigenvectors of $H_{\rm II}^\prime$ given above
	\begin{equation}
		\begin{split}
			\Phi_1^{(\rm II)}=\frac{1}{2\sqrt{v^2k_{\parallel}^2+\frac14(\Delta-\lambda)^2}\big(\sqrt{v^2k_{\parallel}^2+\frac14(\Delta-\lambda)^2}-\frac12(\Delta-\lambda)\big)}\begin{pmatrix}
				v(k_x-\ri k_y)\\
				\sqrt{v^2k_{\parallel}^2+\frac14(\Delta-\lambda)^2}-\frac12(\Delta-\lambda)
			\end{pmatrix}\otimes\begin{pmatrix}
			1\\0
			\end{pmatrix} \\
			\Phi_2^{(\rm II)}=\frac{1}{2\sqrt{v^2k_{\parallel}^2+\frac14(\Delta+\lambda)^2}\big(\sqrt{v^2k_{\parallel}^2+\frac14(\Delta+\lambda)^2}-\frac12(\Delta+\lambda)\big)}\begin{pmatrix}
				v(k_x-\ri k_y)\\
				\sqrt{v^2k_{\parallel}^2+\frac14(\Delta+\lambda)^2}-\frac12(\Delta+\lambda)
			\end{pmatrix}\otimes\begin{pmatrix}
				0\\1
			\end{pmatrix} \\
			\Phi_3^{(\rm II)}=\frac{1}{2\sqrt{v^2k_{\parallel}^2+\frac14(\Delta-\lambda)^2}\big(\sqrt{v^2k_{\parallel}^2+\frac14(\Delta-\lambda)^2}+\frac12(\Delta-\lambda)\big)}\begin{pmatrix}
				v(k_x-\ri k_y)\\
				\sqrt{-v^2k_{\parallel}^2+\frac14(\Delta-\lambda)^2}-\frac12(\Delta-\lambda)
			\end{pmatrix}\otimes\begin{pmatrix}
				1\\0
			\end{pmatrix} \\
			\Phi_4^{(\rm II)}=\frac{1}{2\sqrt{v^2k_{\parallel}^2+\frac14(\Delta+\lambda)^2}\big(\sqrt{v^2k_{\parallel}^2+\frac14(\Delta+\lambda)^2}+\frac12(\Delta+\lambda)\big)}\begin{pmatrix}
				v(k_x-\ri k_y)\\
				\sqrt{-v^2k_{\parallel}^2+\frac14(\Delta+\lambda)^2}-\frac12(\Delta+\lambda)
			\end{pmatrix}\otimes\begin{pmatrix}
				0\\1
			\end{pmatrix}
		\end{split}
	\end{equation}
	\par 
	Notice that $k_{z,1}^{(\rm II)},k_{z,3}^{(\rm II)},\mathcal{E}_1^{(\rm II)},\mathcal{E}_3^{\rm (II)},\Phi_1^{(\rm II)},\Phi_3^{(\rm II)}$ correspond to the $S_z=1$ component, while the remaining quantites correspond to the $S_z=-1$ component.

\section{Density of states calculation}
\label{app:DOS}

	The four bands in region I are \begin{equation}
		E_i(\vec{k})=\mathcal{E}_i^{(\rm I)}+\frac{\hbar^2k_{z,i}^{(\rm I)2}}{2m_0}-\varepsilon_F
	\end{equation}
	Define \begin{equation}
		E_i^\parallel=E_i(\vec k)-\frac{\hbar^2k_{z,i}^{(\rm I)2}}{2m_0}=\mathcal{E}_i^{(\rm I)}-\varepsilon_F
	\end{equation}
	Then \begin{equation}
		E_i(\vec k)=E_i^\parallel+\alpha k_{z,i}^{(\rm I)2}
	\end{equation}
	where $\alpha=\hbar^2/(2m_0)$. Let $X=E-E_i^\parallel$, then the density of states
	\begin{align}
		D_i(E) &=\frac{1}{(2\pi)^3}\iint\di^2\vec{k_\parallel}\int_{-\infty}^{\infty}\di k_z\delta(E-E_i^\parallel-\alpha k_z^2)\\
		       &=\frac{1}{(2\pi)^3}\iint\di^2\vec{k_\parallel}\int_{-\infty}^{\infty}\di k_z\delta(X-\alpha k_z^2)\\
		       &=\frac{1}{(2\pi)^3}\iint\di^2\vec{k_\parallel}\int_{-\infty}^{\infty}\di k_z\frac{1}{2\sqrt{\alpha X}}(\delta(k_z-\sqrt{\frac X\alpha})+\delta(k_z+\sqrt{\frac X\alpha}))
	\end{align}
	The integration \begin{equation}
		\int_{-\infty}^\infty\di k_z\delta(\alpha k_z^2-X)=\begin{cases}
			1/(\sqrt{\alpha X}),X>0\\
			0,X\leq0
		\end{cases}
	\end{equation}
	So that Eq.\ref{eq:DOS} can be turned to\begin{equation}
		D_i(E)=\frac{1}{(2\pi)^3}\int \di^2\vec{k_\parallel}\frac{\Theta(E-E_i^\parallel)}{\sqrt{\frac{\hbar^2}{2m_0}(E-E_i^\parallel)}}=\frac{1}{(2\pi)^2}\int_0^{k_c}k_\parallel\di k_\parallel\frac{\Theta(E-E_i^\parallel)}{\sqrt{\frac{\hbar^2}{2m_0}(E-E_i^\parallel)}}
	\end{equation}
	\par 
	When $i=1,2$, define $E_0=E\pm h_{\rm I}+\varepsilon_F$, $k^\ast=\min(k_c,\frac{E_0}{v_0})$. In practice, the integration range is restricted to $[0,k^\ast]$, where both cutoff condition and the Heaviside function condition are satisfied. Let $\xi=E-E_i^\parallel=E_0-v_0k_\parallel,\di k_\parallel=-\frac{\di \xi}{v_0}$, then\begin{align}
	D_i(E) &=\frac{1}{(2\pi)^2}\int_0^{k^\ast}k_\parallel\di k_\parallel\frac{1}{\sqrt{\frac{\hbar^2}{2m_0}(E_0-v_0k_\parallel)}}\\
	       &=\frac{1}{(2\pi)^2}\int_{E_0}^{E_m}\frac{\di \xi}{v_0}\frac{E_0-\xi}{v_0}\cdot\frac{\sqrt{2m_0}}{\hbar\sqrt{\xi}}\\
	       &=\frac{\sqrt{2m_0}}{(2\pi v_0)^2\hbar}(\frac43 E_0^{\frac32}-2E_0E_m^{\frac12}+\frac23E_m^{\frac32})
	\end{align}
	where $E_m=E_0-v_0k^\ast$.
	\par 
	When $i=3,4$,  define $E_0=E\pm h_{\rm I}+\varepsilon_F$, $k^\ast=\max(0,-\frac{E_0}{v_0})$. The integration range is restricted to $[k^\ast,k_c]$. Let $\xi=E-E_i^\parallel=v_0k_\parallel+E_0$, then \begin{align}
	D_i(E) &=\frac{1}{(2\pi)^2}\int_0^{k^\ast}k_\parallel\di k_\parallel\frac{1}{\sqrt{\frac{\hbar^2}{2m_0}(E_0-v_0k_\parallel)}}\\
		       &=\frac{1}{(2\pi)^2}\int_{E_1}^{E_2}\frac{\di \xi}{v_0}\cdot\frac{\xi-E_0}{v_0}\cdot\frac{\sqrt{2m_0}}{\hbar\sqrt{\xi}}\\
		       &=\frac{\sqrt{2m_0}}{(2\pi v_0)^2\hbar}[\frac23(E_2^{\frac32}-E_1^{\frac32})-2E_0(\sqrt{E_2}-\sqrt{E_1})]
	\end{align}

\section{Condition for interference-induced negative magnetoresistance}
\label{app:proof_inequal}

We consider the inequality
\begin{equation}
\sum_{\lambda=\leftarrow,\rightarrow}
\frac{4k_\lambda^2 k_{\rm II}^2}{\bigl(k_{\rm II}^2+k_\lambda^2\bigr)^2}
\;\le\;
\frac{8k_{\leftarrow}k_{\rightarrow}k_{\rm II}^2}
{\bigl(k_{\rm II}^2+k_{\leftarrow}k_{\rightarrow}\bigr)^2},
\label{ineq-original}
\end{equation}
where
\[
k_{\leftarrow}>0,\qquad k_{\rightarrow}>0,\qquad k_{\rm II}>0.
\]
To determine under what conditions Eq.~\eqref{ineq-original} holds, we introduce the dimensionless variables
\begin{equation}
x=\frac{k_{\leftarrow}}{k_{\rm II}},
\qquad
y=\frac{k_{\rightarrow}}{k_{\rm II}},
\label{xy-def}
\end{equation}
so that Eq.~\eqref{ineq-original} becomes
\begin{equation}
\frac{4x^2}{(1+x^2)^2}
+
\frac{4y^2}{(1+y^2)^2}
\;\le\;
\frac{8xy}{(1+xy)^2}.
\label{ineq-xy}
\end{equation}
Defining
\begin{equation}
\Delta
=
\frac{4x^2}{(1+x^2)^2}
+
\frac{4y^2}{(1+y^2)^2}
-
\frac{8xy}{(1+xy)^2},
\label{Delta-def}
\end{equation}
one finds after straightforward algebra
\begin{equation}
\Delta
=
\frac{
4(x-y)^2
\left[
(xy+1)^2\bigl((xy)^2-4xy+1\bigr)-2xy(x-y)^2
\right]
}
{(1+x^2)^2(1+y^2)^2(1+xy)^2}.
\label{Delta-factorized}
\end{equation}
Since the denominator is strictly positive, Eq.~\eqref{ineq-xy} holds if and only if
\begin{equation}
(xy+1)^2\bigl((xy)^2-4xy+1\bigr)\le 2xy(x-y)^2.
\label{criterion-xy}
\end{equation}
In terms of the original variables, this condition is equivalent to
\begin{equation}
\bigl(k_{\leftarrow}k_{\rightarrow}+k_{\rm II}^2\bigr)^2
\bigl(k_{\leftarrow}^2k_{\rightarrow}^2-4k_{\leftarrow}k_{\rightarrow}k_{\rm II}^2+k_{\rm II}^4\bigr)
\le
2k_{\leftarrow}k_{\rightarrow}k_{\rm II}^2
\bigl(k_{\leftarrow}-k_{\rightarrow}\bigr)^2.
\label{criterion-k}
\end{equation}
Equation~\eqref{criterion-k} therefore gives the necessary and sufficient condition for Eq.~\eqref{ineq-original} to hold.

Notice from Eq. (\ref{eq:expressions_k_FP}) that
\begin{equation}
k_{\rm II}<k_{\leftarrow},\qquad k_{\rm II}<k_{\rightarrow}.
\label{additional-assumption}
\end{equation}
It is convenient to define
\begin{equation}
p=xy=\frac{k_{\leftarrow}k_{\rightarrow}}{k_{\rm II}^2}>1.
\label{p-def}
\end{equation}
Equation~\eqref{criterion-xy} may then be rewritten as
\begin{equation}
(p+1)^2(p^2-4p+1)\le 2p(x-y)^2.
\label{criterion-p}
\end{equation}
If
\begin{equation}
p^2-4p+1\le 0,
\label{p-condition}
\end{equation}
the left-hand side of Eq.~\eqref{criterion-p} is non-positive, while the right-hand side is non-negative, and therefore Eq.~\eqref{criterion-p} is automatically satisfied. Solving Eq.~\eqref{p-condition}, one obtains
\begin{equation}
2-\sqrt{3}\le p\le 2+\sqrt{3}.
\label{p-range}
\end{equation}
Since \(p>1\) under Eq.~\eqref{additional-assumption}, a sufficient condition for Eq.~\eqref{ineq-original} is
\begin{equation}
1<p\le 2+\sqrt{3},
\end{equation}
namely
\begin{equation}
k_{\leftarrow}k_{\rightarrow}\le (2+\sqrt{3})\,k_{\rm II}^2.
\label{sufficient-condition}
\end{equation}

\section{Parameter selection and justification}

	In this appendix, we summarize the physical parameters employed in our calculations. All parameters are chosen based on established literature values or experimentally motivated considerations to ensure consistency with Ref.31. whose experiments were performed in similar system.
	\par 
	The effective Hamiltonian of WSe$_2$ near K valley is modeled using the massive Dirac form with SOC. The material parameters entering the Hamiltonian are taken from the following sources.
	\par 
	Band gap $\Delta$, spin-orbit coupling $\lambda$ and velocity parameter $v$ are adopted from Ref.14 and $m$ is taken from Ref.32. The velocity parameter of graphene is taken from Ref.35. \par 
		The cutoff $k_c$ used in the in-plane momentum integration is fixed by matching the magnitude of the magnetoresistance ratio reported in Ref.32. The spin polarization $P$ extracted from their fitting is used to deduce the effective Fermi-level shift $\varepsilon_F$ in region I and III.
	\end{widetext}

\end{document}